\documentclass[conference]{IEEEtran}
\IEEEoverridecommandlockouts
\usepackage{cite}
\usepackage{amsmath,amssymb,amsfonts}
\usepackage{graphicx}
\usepackage{textcomp}
\usepackage{comment}
\usepackage{float} 

\usepackage{tikz}
\usetikzlibrary{arrows.meta,positioning,shapes.geometric,calc,fit, shadows, backgrounds}

\tikzstyle{block} = [rounded rectangle, minimum width=1cm, minimum height=1cm, text centered, draw=black, thin, inner sep=2pt]
\tikzstyle{arrow} = [thick,->,>=stealth]

\usepackage{color}

\usepackage[utf8]{inputenc} 
\usepackage[T1]{fontenc}    
\usepackage{hyperref}       
\usepackage{url}            
\usepackage{booktabs}       
\usepackage{amsfonts}       
\usepackage{nicefrac}       
\usepackage{microtype}      
\usepackage{xcolor}         
\usepackage{enumitem}
\usepackage{comment}
\usepackage{tcolorbox}
\usepackage{listings}
\usepackage{pdfpages}
\usepackage{longtable}
\usepackage{multirow}
\usepackage{graphicx}
\usepackage{verbatim}
\usepackage{stfloats}
\usepackage[export]{adjustbox}

\usepackage{tabularx}
\usepackage{lipsum} 
\usepackage{colortbl}
\usepackage{float}
\usepackage{hyphenat}

\usepackage{booktabs}
\usepackage{array}
\usepackage{wrapfig}
\usepackage{fancyvrb}

\usepackage{setspace}
\definecolor{swecream}{HTML}{EFFEFF}
\definecolor{issueborder}{HTML}{15071A}
\definecolor{issuefill}{HTML}{F6F8FA}
\definecolor{envfill}{HTML}{F2F9FF}
\definecolor{envborder}{HTML}{123C7C}
\definecolor{agentfill}{HTML}{F9F3F3}
\definecolor{agentborder}{HTML}{9B0A0A}
\definecolor{goldpatchborder}{HTML}{FABB00}
\definecolor{goldpatchfill}{HTML}{FFF7E1}

\DefineVerbatimEnvironment{CodeVerbatim}{Verbatim}{
  formatcom={\color{black}},
  fontsize=\small,
  fontfamily=\ttdefault,
  fontseries=\mddefault,
  fontshape=\updefault,
  fillcolor=\color{white},
  framerule=0pt,
}
\usepackage{longtable}
\newtcolorbox{observationbox}[1][]{
        colback=envfill,
        colbacktitle=envfill,
        colframe=envborder,
        arc=5pt,
        fontupper=\small,
        fonttitle=\bfseries\color{black},
        boxrule=0.5mm,
        boxsep=1mm,
        width=\linewidth,
        breakable,
        title={Observation \hfill #1},
        rounded corners,
        toptitle=0.7mm,
        bottomtitle=0.7mm
}
\newtcolorbox{goldpatchbox}[1][]{
        colback=goldpatchfill,
        colbacktitle=goldpatchfill,
        colframe=goldpatchborder,
        arc=5pt,
        fontupper=\small,
        fonttitle=\bfseries\color{black},
        boxrule=0.5mm,
        boxsep=1mm,
        width=\linewidth,
        breakable,
        title={\twemoji{1f6a9} Flag Captured \hfill #1},
        rounded corners,
        toptitle=0.7mm,
        bottomtitle=0.7mm
}
\newtcolorbox{issuebox}[1][]{
        colback=issuefill,
        colbacktitle=issuefill,
        colframe=issueborder,
        arc=5pt,
        fontupper=\small,
        fonttitle=\bfseries\color{black},
        boxrule=0.5mm,
        boxsep=1mm,
        width=\linewidth,
        breakable,
        title={Issue \hfill #1},
        rounded corners,
        toptitle=1mm
}
\newtcolorbox{agentbox}[1][]{
        colback=agentfill,
        colbacktitle=agentfill,
        colframe=agentborder,
        arc=5pt,
        fontupper=\small,
        fonttitle=\bfseries\color{black},
        boxrule=0.5mm,
        boxsep=1mm,
        width=\linewidth,
        breakable,
        title={EnIGMA \hfill #1},
        rounded corners,
        toptitle=1mm,
        lower separated=false
}
\newtcolorbox{fileviewerbox}[1]{
        enhanced,
        breakable,
        boxrule = 1.5pt,
        fontupper = \small,
        fonttitle = \bf\color{black},
        arc = 5pt,
        rounded corners,
        colframe = black,
        colbacktitle = swecream,
        colback = swecream,
        title = #1,
        left=4pt 
}
\newtcolorbox{promptbox}[1]{
    enhanced,
    breakable,
    boxrule=1pt,  
    fontupper=\small,
    fonttitle=\bfseries\color{black},
    arc=3pt,  
    rounded corners,
    colframe=black,
    colbacktitle=swecream,
    colback=swecream,
    title=#1,
    left=2mm,  
    right=2mm,  
    top=1mm,  
    bottom=1mm  
}

\newcommand{\cmark}{\textcolor{green!60!black}{$\surd$}}
\newcommand{\xmark}{\textcolor{red}{$\times$}} 
\newcommand{\pmark}{\textcolor{orange!85!black}{$\triangle$}}
\newcommand{\rev}[1]{#1}
\newcounter{algctr}
\makeatletter
\newcommand{\SetAlgoNoEnd}{}
\newcommand{\KwIn}[1]{\par\noindent\textbf{Input:} #1\par}
\newcommand{\KwOut}[1]{\par\noindent\textbf{Output:} #1\par}

\newcommand{\algindent}[1]{\par\begingroup\leftskip=1em #1\par\endgroup}

\newcommand{\ForEach}[2]{\par\noindent\textbf{for each} #1 \textbf{do}\algindent{#2}}
\newcommand{\If}[2]{\par\noindent\textbf{if} #1 \textbf{then}\algindent{#2}}
\newcommand{\Else}[1]{\par\noindent\textbf{else}\algindent{#1}}
\newcommand{\Return}[1]{\par\noindent\textbf{return} #1\par}
\newenvironment{algorithm}[1][]{%
  \refstepcounter{algctr}%
  \begin{tcolorbox}[colback=white,colframe=black!60,boxrule=0.45pt,arc=1pt,
  left=3pt,right=3pt,top=3pt,bottom=3pt]%
  \renewcommand{\caption}[1]{\par\noindent\textbf{Algorithm~\thealgctr: ##1}\par\vspace{2pt}}%
  \let\oldsemicolon\;%
  \renewcommand{\;}{\par}%
  \raggedright
}{%
  \end{tcolorbox}
}
\makeatother

\def\BibTeX{{\rm B\kern-.05em{\sc i\kern-.025em b}\kern-.08em
    T\kern-.1667em\lower.7ex\hbox{E}\kern-.125emX}}
\begin{document}
\bstctlcite{IEEEexample:BSTcontrol}

\title{EEG-Fusion: Failure-Informed Source-Free Expert Routing for Robust Motor Imagery EEG Decoding}

\author{
\IEEEauthorblockN{Abdul Basit, \quad Saim Rehman, \quad Muhammad Shafique}
\IEEEauthorblockA{\textit{eBRAIN Lab, Division of Engineering} \textit{New York University (NYU) Abu Dhabi}, Abu Dhabi, UAE\\
abdul.basit@nyu.edu, \quad sr7849@nyu.edu, \quad muhammad.shafique@nyu.edu}
\vspace{-20pt}
}

\maketitle

\begin{abstract}
Subject-independent motor-imagery (MI) EEG decoding can \rev{fail at subject level despite acceptable average performance}: under subject shift, a decoder can become an overconfident near-one-class predictor. \rev{In source-free deployment, target labels are unavailable for adaptation or selection.} We present \textit{EEG-Fusion}, a decision-level framework \rev{that predicts candidate quality, observes label-free collapse, and routes accordingly}. \rev{Three neural architectures undergo Euclidean alignment and normalization-only adaptation. A source-held-out gate estimates performance from unlabeled diagnostics and penalizes collapsed outputs.} Diagnostics include confidence, entropy, diversity, agreement, and class balance; collapse is \rev{the maximum predicted class fraction}. In 9-fold leave-one-subject-out (LOSO) evaluation with three seeds, relative to \rev{NoAlign EEGNet}, \rev{mean macro-F1 increases} from 0.417 to \rev{0.524} on BCI IV-2a local, 0.314 to \rev{0.480} on BNCI2014-001, and 0.607 to 0.708 on BNCI2014-004; collapse reductions are \rev{0.194, 0.224}, and 0.169. \rev{The three $n=9$ macro-F1 pipeline-versus-anchor tests do not pass Holm correction. On all 52 Cho2017 subjects, F1 increases from 0.629 to 0.709 and collapse decreases from 0.719 to 0.536. Relative to EA+Safe-BN EEGNet, routing preserves F1 ($+0.0004$, adjusted $p=0.755$) while reducing collapse by 0.014 (adjusted $p=0.0026$). Alignment/adaptation therefore provides most accuracy gain; routing supplies a label-free reliability refinement.}
\end{abstract}

\begin{IEEEkeywords}
Motor imagery EEG, source-free adaptation, test-time adaptation, expert routing, class collapse, BCI.
\end{IEEEkeywords}

\section{Introduction}
\label{sec:introduction}

Motor-imagery (MI) brain--computer interfaces (BCIs) can support neurorehabilitation and assistive control without invasive sensing, but their utility depends on whether a decoder trained on prior users remains reliable for a new user \cite{nicolas2012brain,daly2008brain}. This is difficult because MI-EEG is low-SNR, non-stationary, and strongly subject dependent \cite{lotte2018review,tangermann2012review}. We focus on source-free, calibration-free batch deployment: after source-side training, the system receives an unlabeled target EEG stream for alignment, adaptation, diagnostics, and routing; target labels are accessed only after predictions are fixed for evaluation.

\begin{figure}[t]
\centering
\includegraphics[width=\linewidth]{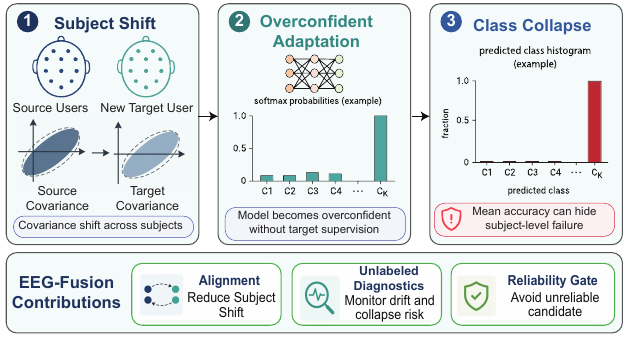}
\caption{Motivation for EEG-Fusion. Under subject shift, source-free MI-EEG decoders can become confident but collapsed. EEG-Fusion treats this as a label-free reliability problem using stream diagnostics and collapse-aware expert routing.}
\label{fig:intro_motivation}
\end{figure}

Under this constraint, a model may look acceptable on average while producing a collapsed prediction distribution for a particular held-out subject. Post-hoc analysis revealed a recurring failure mode that we call \textit{class collapse}: the predicted class histogram concentrates on one MI class despite a balanced target protocol. We summarize this behavior by the class-collapse index, the maximum predicted class fraction in the target stream. \rev{Collapse is read directly from the unlabeled prediction histogram, as summarized in Fig.~\ref{fig:intro_motivation}.}

The literature provides strong components but not a complete answer to this deployment problem. Classical approaches such as FBCSP and Riemannian covariance classifiers are data-efficient and physiologically grounded \cite{4634130,6046114}, while compact deep models such as EEGNet and temporal/filter-bank CNNs learn useful end-to-end representations \cite{lawhern2018eegnet,schirrmeister2017deep,mane2021fbcnet,salami2022eegitnet}. Recent work has further advanced cross-subject MI decoding through domain generalization, online test-time adaptation, and EEG foundation models \cite{zhong2023eegdg,wimpff2023otta,li2024ttime,jiang2024labram,wang2024cbramod,liu2025mirepnet}. These methods improve representations or adaptation, but they are typically assessed primarily through aggregate accuracy or macro-F1. Table~\ref{tab:related_positioning} summarizes the resulting gap: label-free subject-level collapse diagnosis and reliability-controlled expert selection remain comparatively underexplored.

\rev{Preliminary raw-plus-handcrafted feature concatenation was not consistently beneficial. We therefore retain physiological and covariance models as independent comparators, while the primary router asks which of three adapted neural architectures is reliable for each new subject.}

To address this problem, we propose \textit{EEG-Fusion}, a failure-informed reliability-modeling framework for source-free MI decoding. \rev{It runs three adapted decoders, predicts their likely performance from unlabeled target diagnostics, penalizes observed collapse, and selects the highest score. \emph{Fusion} denotes this decision-level selection. A routed candidate is EEGNet, SimpleConv, or a multi-kernel temporal CNN after the same safety BN-TTA; the policy study treats architecture--adaptation pairs as distinct candidates.} The novel contributions are:
\begin{itemize}[leftmargin=*]
    \item \textbf{Failure-aware source-free formulation:} We explicitly measure class collapse and subject-level performance variation, rather than relying only on mean accuracy.
    \item \textbf{Reliability-gated expert routing:} We train a source-held-out gate to \rev{predict candidate performance and combine it with observed label-free collapse}, and deploy it on unseen target subjects without target labels. \rev{A controlled extension also tests architecture--adaptation-policy candidates.}
    \item \textbf{Auditable adaptation protocol:} We combine subject-wise Euclidean alignment with \rev{parameter-scoped} normalization-only test-time adaptation and log pre/post-adaptation diagnostics, selected states, and routing decisions before target labels are accessed.
    \item \textbf{Multi-benchmark LOSO evaluation:} We compare \rev{routed neural architectures with} filter-bank, covariance-based, \rev{and} physiological-feature \rev{comparators} under leave-one-subject-out (LOSO) with paired tests and bootstrap \rev{CIs}.
\end{itemize}

\begin{table*}[b]
\centering
\scriptsize
\caption{Positioning versus representative MI-EEG/calibration-free BCI methods. \cmark\ central, \pmark\ partial, \xmark\ not primary.}
\label{tab:related_positioning}
\setlength{\tabcolsep}{3.1pt}
\resizebox{\textwidth}{!}{%
\begin{tabular}{p{2.2cm}p{5.5cm}ccccc}
\toprule
\textbf{Method family} & \textbf{Representative works} & \textbf{\rev{Unlabeled target adapt.}} & \textbf{Expert routing} & \textbf{Collapse-aware} & \textbf{\rev{Classical baselines}} & \textbf{Pre-label audit} \\
\midrule
Spectral-spatial and Riemannian baselines & FBCSP, CSP/LDA, tangent-space and MDM classifiers \cite{4634130,6046114} & \pmark & \xmark & \xmark & \cmark & \xmark \\
Compact deep MI decoders & EEGNet, DeepConvNet, EEG-TCNet, FBCNet, EEG-Inception, EEG-ITNet \cite{lawhern2018eegnet,schirrmeister2017deep,ingolfsson2020eeg_tcnet,mane2021fbcnet,zhang2021eeg_inception,salami2022eegitnet} & \xmark & \xmark & \xmark & \xmark & \xmark \\
Domain generalization & Multi-source invariance and unseen-subject training, e.g., EEG-DG \cite{zhong2023eegdg} & \xmark & \xmark & \xmark & \xmark & \xmark \\
Online/source-free TTA & Adaptive BN, entropy minimization, information maximization, SPDIM, and T-TIME \cite{wang2021tent,wimpff2023otta,li2024spdim,li2024ttime} & \cmark & \rev{\pmark} & \pmark & \xmark & \pmark \\
Pretrained EEG representation models & LaBraM, CBraMod, MIRepNet \cite{jiang2024labram,wang2024cbramod,liu2025mirepnet} & \xmark & \xmark & \xmark & \xmark & \xmark \\
\textbf{EEG-Fusion} & \rev{Source-free reliability routing over adapted neural candidates; classical comparators} & \cmark & \cmark & \cmark & \cmark & \cmark \\
\bottomrule
\end{tabular}}
\end{table*}

Across the completed LOSO protocols, EEG-Fusion \rev{yields higher mean subject macro-F1 and lower class collapse than} the \rev{NoAlign EEGNet} source-free anchor, with F1 differences of \rev{0.080--0.166}. \rev{Full 52-subject Cho2017 evaluation provides the primary external dataset check. The intended scope is a reliability-management layer above adaptation: predict candidate quality, observe collapse, and route without target labels.}

\section{Related Work}
\label{sec:related}

\textbf{MI-EEG decoders and reproducible benchmarks:}
Classical MI-EEG pipelines use spectral-spatial filters and shallow classifiers, with FBCSP and Riemannian covariance methods remaining important data-efficient and interpretable baselines \cite{4634130,6046114}. Deep decoders such as DeepConvNet, EEGNet, EEG-TCNet, FBCNet, EEG-Inception, and EEG-ITNet learn temporal, spatial, and filter-bank representations \cite{schirrmeister2017deep,lawhern2018eegnet,ingolfsson2020eeg_tcnet,mane2021fbcnet,zhang2021eeg_inception,salami2022eegitnet}. MOABB and Braindecode studies show that rankings depend strongly on dataset, preprocessing, and protocol details \cite{jayaram2018moabb,chevallier2024moabb}. Because these families differ in inductive bias, data efficiency, and shift sensitivity, EEG-Fusion \rev{routes among compact neural architectures while retaining classical families as comparators} rather than committing to one backbone.
EEG-controlled prosthetic systems further motivate reliable target-user operation beyond offline accuracy \cite{Basit_2025,CognitiveArm}.

\textbf{Cross-subject generalization and source-free adaptation:}
Transfer learning and Euclidean alignment reduce inter-subject covariance shift without target labels \cite{he2018euclidean}. Domain generalization methods such as EEG-DG target unseen-subject invariance \cite{zhong2023eegdg}, while general Test-Time Adaptation (TTA) methods such as TENT and EEG-specific online or source-free methods use normalization updates, entropy minimization, marginal-diversity terms, or information maximization during inference \cite{wang2021tent,wimpff2023otta,li2024ttime,li2024spdim}. \rev{Existing TTA/ensemble methods primarily optimize or combine target-time predictions. EEG-Fusion instead learns source-held-out subject-level quality estimates to select among adaptation candidates.} Inspired by calibration and selective prediction \cite{guo2017calibration,geifman2017selective}, EEG-Fusion treats adaptation as one component of a source-free reliability problem: \rev{predict candidate quality, observe collapse, and route from unlabeled stream diagnostics}.

\textbf{Foundation models and the novelty boundary:}
Large EEG representation models such as LaBraM, CBraMod, and MIRepNet suggest that the strongest raw-EEG backbone may change as pretraining data and architectures evolve \cite{jiang2024labram,wang2024cbramod,liu2025mirepnet}. EEG-Fusion is therefore not another raw-EEG architecture, it is a deployment-level reliability layer \rev{tested here over compact neural candidates}. \rev{Adding a pretrained model requires its own source-held-out reliability records and controlled channel/preprocessing evaluation}, because source-free deployment still requires a label-free mechanism for deciding whether the current subject's predictions are reliable.

\section{Methodology}
\label{sec:method}

\begin{figure*}[t]
\centering
\includegraphics[width=\textwidth]{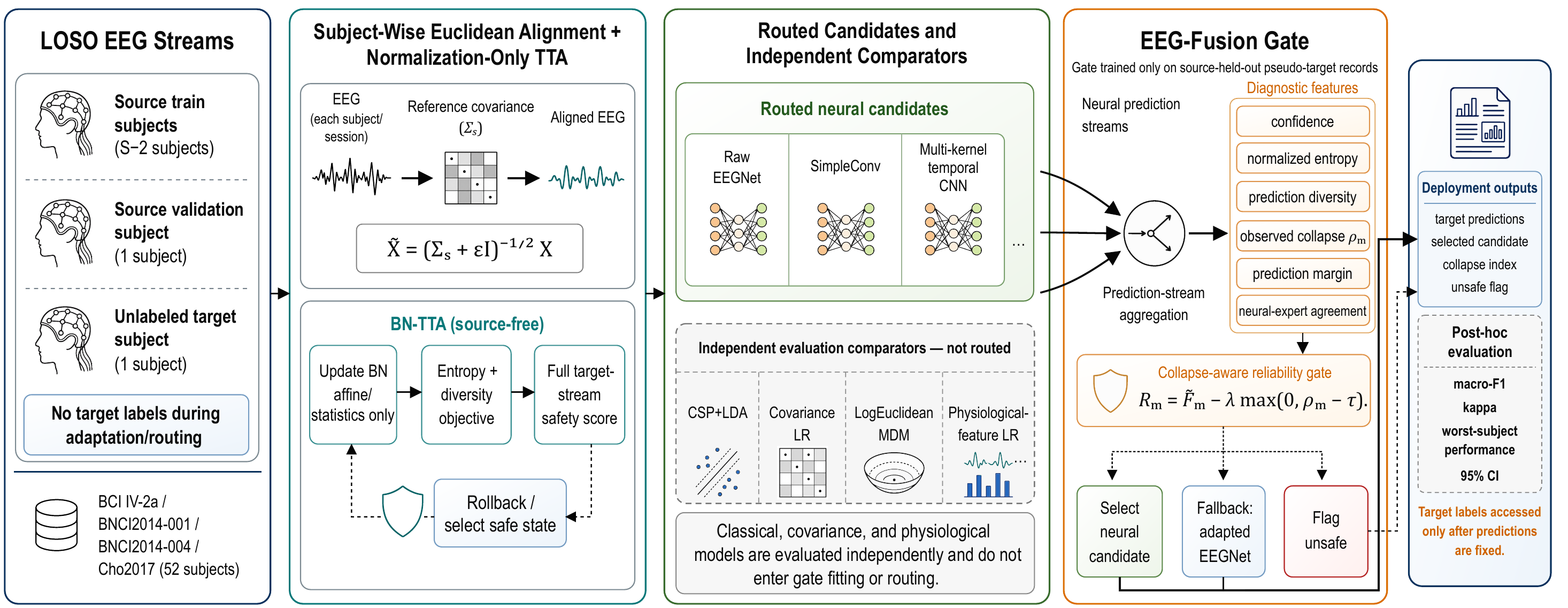}
\caption{\rev{Overview of EEG-Fusion. The source-held-out gate uses unlabeled target diagnostics to select among three normalization-adapted neural candidates, with an adapted raw-input EEGNet fallback. Classical/covariance/physiological models are independent comparators. Target labels are accessed only after predictions are fixed.}}
\label{fig:method_overview}
\end{figure*}

\subsection{Problem Setup}
For an EEG trial $X\in\mathbb{R}^{C\times T}$ and label $y\in\{1,\ldots,K\}$, LOSO decoding trains on source subjects $\mathcal{S}_{src}$ and evaluates on an unseen target subject $\mathcal{S}_{tgt}$. We use \emph{source-free} for target time: after source-side \rev{candidates} and reliability gates are prepared, EEG-Fusion uses trained components and the unlabeled target stream only, with no target labels or source data/labels. The protocol is batch-transductive: the unlabeled target stream is available before final scoring for alignment, adaptation, diagnostics, and routing. Our evaluation objective is to preserve subject-level macro-F1 while reducing degenerate prediction distributions; at target time this is approximated using label-free diagnostics and source-held-out reliability estimates.

We define class-collapse index as
\begin{equation}
    \rho(\hat{Y})=\max_{k\in\{1,\ldots,K\}}\frac{1}{N}\sum_{i=1}^{N}\mathbb{1}[\hat{y}_i=k],
\end{equation}
where $\rho=1$ indicates a one-class predictor and lower values indicate more diverse predictions. Collapse is not a surrogate for accuracy, but it is a useful target-stream safety diagnostic when labels are unavailable.
For a balanced prediction histogram, $\rho\approx1/K$; we compute $\rho_{\mathrm{norm}}=(\rho-1/K)/(1-1/K)$ for cross-task interpretation. \rev{Gates are fitted separately within each fixed-$K$ protocol. Final runs use common raw-$\rho$ guardrails $\tau_s=0.60$ and $\kappa=0.85$, whose normalized values are $(0.47,0.80)$ for four-class and $(0.20,0.70)$ for binary tasks.} \rev{For unknown or naturally imbalanced priors, $\rho$ is only a concentration alert, not evidence of error.} The full pipeline is summarized in Fig.~\ref{fig:method_overview}.

\subsection{Alignment-First Source-Free Preprocessing}
Each subject/session is first transformed by unsupervised Euclidean alignment \cite{he2018euclidean}. For a channel-demeaned trial, let
\begin{equation}
    G(X)=\frac{XX^{\top}}{T-1},\qquad
    \bar G(X)=\frac{G(X)}{\operatorname{tr}(G(X))+\epsilon_c}C .
\end{equation}
The subject/session reference covariance is
\begin{equation}
    \Sigma_s=(1-\gamma)\frac{1}{N_s}\sum_i\bar G(X_{s,i})
    +\gamma\frac{\operatorname{tr}(\bar G_s)}{C}I ,
\end{equation}
where $\gamma$ is a small identity shrinkage coefficient and $\bar G_s=N_s^{-1}\sum_i\bar G(X_{s,i})$. Trials are aligned as
\begin{equation}
    \tilde{X}_{s,i}=(\Sigma_s+\epsilon I)^{-1/2}X_{s,i},
\end{equation}
using only trials from the same subject/session and no labels; the inverse square root is computed by eigenvalue decomposition with the same floor $\epsilon$. Alignment is computed independently for source and target streams. Final runs use trace normalization, $\gamma=10^{-3}$, and $\epsilon_c=\epsilon=10^{-6}$, all recorded in run manifests.

\subsection{\rev{Candidate and Comparator Bank}}
\rev{EEG-Fusion maintains a neural candidate set alongside independent classical comparators. The primary set contains raw-input EEGNet, SimpleConv, and a compact multi-kernel temporal CNN, each after safety BN-TTA.} \rev{Classical comparators} include CSP+LDA, covariance logistic regression, physiological-feature logistic regression, physiological+covariance logistic regression, and LogEuclidean MDM. \rev{They are independent comparators, not routed candidates or gate inputs.} A feature-gated neural model is evaluated as a diagnostic baseline, but EEG-Fusion does not depend on it.

Conventional features are retained as \rev{comparator} inputs, not as a standalone fusion claim. Final runs use time-domain statistics, Hjorth parameters, Welch bandpower/spectral summaries over theta 4--8 Hz, alpha 8--13 Hz, mu 8--12 Hz, beta 13--30 Hz, and low-gamma 30--40 Hz, autoregressive coefficients, and hemispheric asymmetry; entropy, fractal, and STFT-texture families are ablation-only options. \rev{Covariance comparators} use regularized log-covariance vectors or LogEuclidean distances. Confidence diagnostics use native \rev{neural-candidate} probabilities; model identity allows the gate to learn probability-scale differences indirectly.

\subsection{Normalization-Only Test-Time Adaptation}
For neural \rev{candidates}, EEG-Fusion performs source-free test-time adaptation (TTA) only on a restricted parameter set: \rev{batch-normalization affine parameters and running statistics} for compact \rev{candidates} and router/gate parameters for the feature-gated diagnostic baseline. We use ``parameter-scoped'' to describe the update scope, not to claim a measured runtime advantage; full backbone updates are avoided. For an unlabeled target mini-batch of size $B$, adaptation minimizes
\begin{equation}
    \mathcal{L}_{tta}=
    \alpha\frac{1}{B}\sum_{i=1}^{B}H(p_i)
    +\beta\sum_{k=1}^{K}\bar{p}_k\log(\bar{p}_k+\epsilon)
    +\eta\mathcal{R}_{gate},
\end{equation}
where $p_i=\mathrm{softmax}(z_i)$, $\bar{p}_k=B^{-1}\sum_i p_{i,k}$, $H(\cdot)$ is conditional entropy, and the second term is negative marginal entropy, so minimizing it encourages class diversity. For the feature-gated diagnostic baseline only, channel gates $a_{i,c}$ are regularized by $\mathcal{R}_{gate}=(BC)^{-1}\sum_{i,c}|a_{i,c}-1|$; it is zero for non-gated \rev{candidates}.

State selection uses the full unlabeled target stream. Define
\begin{equation}
    \begin{aligned}
    q_k&=N^{-1}\sum_i\mathbb{1}[\hat y_i=k],\\
    D(\hat{Y})&=-\frac{1}{\log K}\sum_k q_k\log(q_k+\epsilon),\\
    \bar H&=\frac{1}{N\log K}\sum_i H(p_i).
    \end{aligned}
\end{equation}
We score a TTA state by
\begin{equation}
    \begin{split}
    S(d)=&D(\hat{Y})+c(1-\bar H)-\mu\max(0,\rho-\tau_s)\\
    &-\nu\max(0,D_{\min}-D(\hat{Y})).
    \end{split}
\end{equation}
The confidence term is intentionally weak ($c=0.25$) and is used jointly with class-histogram entropy and collapse penalties, so confident states are favored only when prediction diversity remains acceptable. The initial, best, and final states are logged. After TTA, the highest-scoring full-stream state is restored; if raw $\rho$ exceeds a protocol-local threshold, training stops early and rolls back to the best logged state. Final neural runs use $Q=10$, $\alpha=1$, $\beta=0.5$, $c=0.25$, $\mu=2$, $\nu=1$, $\tau_s=0.60$, $D_{\min}=0.70$, and collapse threshold $0.85$; $\eta=0.01$ only for the feature-gated baseline. All TTA and routing hyperparameters are fixed a priori or from source-held-out records only.

\subsection{Failure-Informed Reliability Gate}
\rev{For each candidate, the gate predicts candidate macro-F1 from unlabeled diagnostics, observes collapse directly, and selects the best predicted quality after a collapse penalty.}

For target fold $t$, reliability records $\mathcal{H}_t$ come from source-held-out pseudo-target folds in the same protocol. Each source subject $h\neq t$ is treated as a pseudo-target: \rev{candidate models} are trained on $\mathcal{S}_{src}\setminus\{h\}$ and diagnostics $d_{m,h}$ are computed without labels. \rev{Pseudo-target labels are used only to compute the quality target $F_{m,h}$; $\rho_{m,h}$ is observed label-free from the prediction histogram.} A protocol with $S$ subjects and $|\mathcal{M}|$ \rev{candidates} yields $(S-1)|\mathcal{M}|$ records per seed before pooling. \rev{Fold/seed records remain separate; all target-$t$ records are excluded from gate fitting and hyperparameter selection.}

For each target fold, the gate \rev{fits one ridge regressor} with fixed $\lambda_{\mathrm{ridge}}=1$ for $\widehat{F}_{m}$ using source-standardized diagnostics and one-hot model identity; predictions are clipped to $[0,1]$. \rev{A subject-ID assertion excludes target $t$ across all seeds.} Diagnostics include confidence, entropy, margin, class fractions, observed collapse, diversity, confidence rank/gap, and \rev{cross-candidate agreement}. No target-label calibration is performed; model identity captures source-estimated probability-scale differences. \rev{Candidates} are ranked by
\begin{equation}
    \rev{R_m=\widehat{F}_{m}-\lambda\max(0,\rho_m-\tau)},
\end{equation}
where $\rho_m$ is \rev{directly observed label-free collapse}. The top-ranked \rev{candidate} $m^\dagger$ is selected unless $\rho_{m^\dagger}\geq\kappa$, which triggers the fixed fallback $m_0$: \rev{the aligned raw-input EEGNet after safety BN-TTA and best-state restoration}. Final runs use $\lambda=0.25$, $\tau=0.60$, and $\kappa=0.85$. 
\rev{Collapse is used directly as a label-free routing signal alongside predicted candidate quality.} Algorithm~\ref{alg:eeg_fusion_gate} formalizes \rev{the reliability-based} selection. \rev{The same reliability interface is further evaluated with matched-budget TENT architecture--policy candidates.}

\begin{algorithm}[ht]
\caption{Source-Held-Out Reliability Selection}
\label{alg:eeg_fusion_gate}
\SetAlgoNoEnd
\footnotesize
\KwIn{Held-out source records \rev{$\mathcal{H}=\{(d_{m,h},F_{m,h})\}$}, target diagnostics $\{d_{m,t}\}_{m\in\mathcal{M}}$, $\tau,\lambda,\kappa$, fallback $m_0$; \rev{$\mathcal{M}$ contains three neural candidates}}
\KwOut{Selected \rev{candidate} $m^\star$ and reliability scores $\{R_m\}_{m\in\mathcal{M}}$}
Compute standardization statistics from continuous diagnostics in $\mathcal{H}$ and standardize source-held-out and target diagnostics\;
\rev{Append one-hot candidate identity and fit ridge regressor $g_F:d\mapsto\widehat{F}$ using $\mathcal{H}$}\;
\ForEach{\rev{candidate} $m\in\mathcal{M}$}{
    \rev{$\widehat{F}_{m,t}\leftarrow\operatorname{clip}_{[0,1]}g_F(d_{m,t})$; $\rho_{m,t}\leftarrow\max_k N^{-1}\sum_i\mathbb{1}[\hat y_{m,i}=k]$}\;
    \rev{$R_m\leftarrow\widehat{F}_{m,t}-\lambda\max(0,\rho_{m,t}-\tau)$}\;
}
$m^\dagger\leftarrow\arg\max_{m\in\mathcal{M}}R_m$\;
\If{\rev{$\rho_{m^\dagger,t}\geq\kappa$} and $m_0\in\mathcal{M}$}{
    $m^\star\leftarrow m_0$\;
}
\Else{
    $m^\star\leftarrow m^\dagger$\;
}
\rev{Log $\{\widehat{F}_{m,t},\rho_{m,t},R_m\}_{m\in\mathcal{M}}$ before target labels are used}\;
\Return{$m^\star,\{R_m\}_{m\in\mathcal{M}}$}\;
\end{algorithm}

\subsection{Auditability and Label Hygiene}
Target labels are excluded from every adaptation and routing input and enter only after predictions are written for subject-level metrics and paired tests. Each fold/seed/model produces a manifest, pre/post TTA predictions, adaptation diagnostics, \rev{model records}, EEG-Fusion choices, fold metrics, and aggregate statistical tables. This audit trail separates adaptation, routing, and post-hoc evaluation. 

\section{Experimental Setup}
\label{sec:experiment}

\subsection{Datasets and Protocol}
All experiments use subject-independent leave-one-subject-out (LOSO) evaluation. In fold $f$, the target is the $f$th subject in dataset order, validation is the next subject cyclically, and all remaining subjects train the \rev{candidate model}. This source-only rule is fixed across models/seeds. For source-held-out reliability records, pseudo-target $h$ is excluded from training/checkpointing, and validation is chosen cyclically from the remaining source subjects. \rev{Gate fitting excludes every record of the target subject across seeds, enforced by a subject-ID disjointness assertion.} Target labels are used only after predictions are fixed. Table~\ref{tab:datasets} summarizes the completed protocols \cite{tangermann2012review,jayaram2018moabb,chevallier2024moabb,cho2017eeg}. Local BCI IV-2a uses T-session GDF trials, BNCI2014-001 uses MOABB IV-2a with both sessions, and \rev{Cho2017 uses all 52 MOABB subjects}. Because local BCI IV-2a and BNCI2014-001 share the same IV-2a task family but differ in loader, session usage, and preprocessing, we interpret them as a protocol/reproducibility check rather than independent biological evidence.

\begin{table}[ht]
\centering
\scriptsize
\caption{Completed protocols. \rev{Core protocols use 9 LOSO folds and Cho2017 uses 52}; all use 4--40 Hz signals at 128 Hz, 50 neural epochs, and 10-step BN-TTA. Trials are representative train/validation/test counts.}
\label{tab:datasets}
\setlength{\tabcolsep}{2.7pt}
\resizebox{\columnwidth}{!}{%
\begin{tabular}{lccccc}
\toprule
\textbf{Dataset} & \textbf{Use} & \textbf{Task} & \textbf{Ch.} & \textbf{\rev{Folds}} & \textbf{Trials/fold} \\
\midrule
BCI IV-2a local & T only & 4-class MI & 22 & \rev{9} & 2016/288/288 \\
BNCI2014-001 & MOABB T+E & 4-class MI & 22 & \rev{9} & 4032/576/576 \\
BNCI2014-004 & MOABB all & left/right MI & 3 & \rev{9} & 5120/680/720 \\
\rev{Cho2017 full} & \rev{MOABB all} & left/right MI & 64 & \rev{52} & \rev{$\sim$10115/202/202} \\
\bottomrule
\end{tabular}}
\end{table}

Signals are resampled to 128 Hz and restricted to 4--40 Hz. The local IV-2a path uses 22 EEG channels, 50 Hz notch, zero-phase MNE FIR band-pass, average reference, and a 0.5--3.5 s cue-locked MI window without baseline correction. Completed local runs used automatic per-subject/session FastICA/EOG correction ($n_{\mathrm{components}}=0.99$, random state 42, MNE EOG-correlation rejection); failures abort rather than exclude subjects silently. MOABB code sets only $f_{\min}=4$, $f_{\max}=40$, and resample 128; epoch/reference behavior follows dataset-specific MOABB defaults under the logged package version, with no added notch/ICA. Resolved sample counts, package versions, and environment details are logged in manifests. Euclidean alignment is applied independently per subject/session.

\subsection{Baselines and EEG-Fusion Configurations}
We report six groups: \rev{NoAlign EEGNet} anchor, aligned neural \rev{candidates} before/after BN-TTA, \rev{classical comparators}, a feature-gated diagnostic baseline, and EEG-Fusion routing. \rev{Raw-input EEGNet} uses $F_1=16,D=2$ and kernel length 64; SimpleConv uses 32/64-filter Conv1D-BN-ELU-pooling blocks with kernels 15/9; the multi-kernel temporal CNN, reported as FilterBank CNN in result logs, uses temporal kernels $(15,31,63)$ and depthwise spatial filtering on the same 4--40 Hz input, not separate sub-bands. \rev{These three adapted architectures form the routed candidate set; the fallback is EA+Safe-BN EEGNet after best-state restoration. Classical comparators are} CSP+LDA, \rev{covariance and physiological-feature logistic regression, their combined variant,} and LogEuclidean MDM. \rev{All classical and feature-gated models are independent comparators.}

Neural models use seeds $\{11,22,33\}$, 50 epochs, AdamW, learning rate $10^{-3}$, weight decay $10^{-4}$, batch size 64, dropout 0.35, cross-entropy, and validation macro-F1 checkpoints. No early stopping, class balancing, or augmentation is used because the selected MI protocols are class-balanced by design. BN-TTA uses the full target stream for 10 steps, batch size 64, learning rate $5\!\times\!10^{-4}$, and updates only \rev{BN affine parameters and running statistics} per fold/seed/\rev{candidate}; the feature-gated baseline updates router+BN parameters. \rev{Matched-budget, not independently optimized, TENT reloads the aligned checkpoint per subject, updates BN affine parameters with running statistics disabled for $Q=10$ unshuffled steps, and resets episodically.} Classical defaults are fixed: MNE CSP uses six Ledoit--Wolf log-variance components with MNE's default multiclass CSP implementation under the logged package version and LDA. Logistic regression uses train-fit standardization, L2 penalty, $C=1$, lbfgs solver, balanced weights, and 3000 iterations. Covariance logistic regression uses shrinkage-0.1 log-covariance upper-triangular features; MDM uses shrinkage-0.1 LogEuclidean centroids. Classical scripts are indexed on the same seed grid for pairing; deterministic repeats are not additional subject observations because final tests average seeds first. For each target fold, EEG-Fusion fits the gate using only pseudo-target records that do not involve the true target subject. The 9-fold, 3-seed protocols yield 27 fold/seed runs per method, averaged to 9 held-out-subject observations for testing; \rev{full Cho2017 yields 156 runs averaged to 52 subjects}.

\subsection{Reproducibility Artifacts}
Every run writes a manifest with dataset/session, target/validation IDs, split sizes, channels, window/filter settings, seed, optimizer/TTA settings, code hashes, package environment, checkpoint, fallback \rev{candidate}, and selected EEG-Fusion \rev{candidate}. Pre/post-TTA predictions, reliability records, \rev{predicted macro-F1, observed collapse}, risk scores, and fold metrics generate all tables and figures.

\rev{An uncontended GPU profile over three IV-2a folds (288 trials, seed 11; loading excluded) measures $4.66\pm0.30$ s ($16.2\pm1.1$ ms/trial) across three candidates: $0.242\pm0.079$ s alignment, $0.293\pm0.016$ s pre-TTA inference, and $4.42\pm0.24$ s Q10 BN-TTA. Peak allocation/model storage are 520/0.14 MB; routing overhead is negligible.}

\subsection{Metrics and Statistics}
Primary metrics are per-subject macro-F1 and raw class-collapse index $\rho$; accuracy, kappa, and $\rho_{\mathrm{norm}}$ are audit metrics. Pairwise comparisons use seed-averaged subjects, 10,000 paired bootstraps, two-sided Wilcoxon tests, and exact sign flips for $n=9$. \rev{Holm-adjusted Wilcoxon tests support primary inference, while sign flips are corroborative and bootstrap CIs describe mean effects. Protocol/metric families contain pipeline-vs-anchor and routing-vs-adapted; safety-vs-TENT is secondary and paired. Identity-quality replaces the diagnostic F1 predictor with candidate identity while retaining observed-collapse control. Its quality-only variant also disables the collapse penalty and fallback. For source-pool stability, each target/pool size (2/4/6/8) uses up to 20 fixed-seed subject combinations sampled without replacement. Regret averages three model seeds (180 draws per $n=9$ protocol; 1040 for Cho52). Oracle regret $\max_m F_{m,t}-F_{m^\star,t}$ is computed only after label-free selection.} \rev{Prior stress averages 10 fixed-seed post-TTA subsamples at 0.70/0.85 majority fractions.}

\section{Results and Discussion}
\label{sec:results}

\subsection{Main LOSO Results}
\begin{figure}[ht]
\centering
\includegraphics[width=\linewidth]{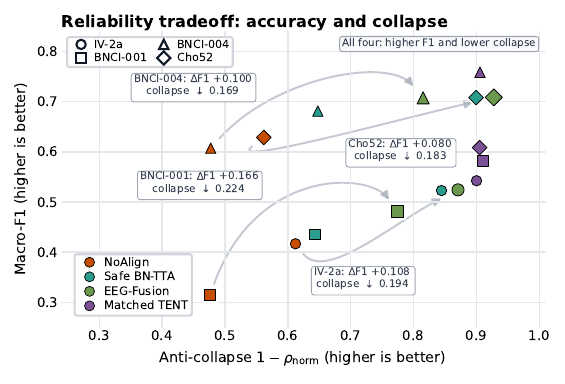}
\caption{\rev{Accuracy--reliability tradeoff. Arrows connect NoAlign EEGNet to EEG-Fusion; callouts report $\Delta$F1 and raw-collapse reduction.}}
\label{fig:rf1_tradeoff}
\end{figure}

Table~\ref{tab:main_results} \rev{separates full-pipeline gains from routing-specific effects and includes the matched-budget TENT (Q10) baseline; the post-hoc classical results are reported for contextual comparison. Alignment and adaptation account for most of the improvement over the NoAlign anchor, and the three $n=9$ macro-F1 pipeline-versus-anchor comparisons do not remain significant after Holm correction. On full Cho2017, routing over EA+Safe-BN EEGNet produces a negligible F1 change ($+0.0004$, adjusted $p=0.755$) while reducing collapse by 0.014 [0.0067, 0.0223] (adjusted $p=0.0026$). Relative to NoAlign EEGNet, EEG-Fusion increases F1 by 0.080 [0.051, 0.110] and reduces collapse by 0.183 [0.149, 0.216]. Fig.~\ref{fig:rf1_tradeoff} illustrates this accuracy--reliability tradeoff.}

\begin{table*}[t]
\centering
\scriptsize
\caption{\rev{Subject results (F1/C; C: collapse; higher F1/lower C are better). TENT uses Q10; deltas use unrounded subject means. CIs are bootstrap mean effects; Holm-Wilcoxon controls inference (no $n=9$ macro-F1 pipeline-vs-anchor test is significant). Post-hoc Classical$^\dagger$: MDM/IV-2a, covariance LR/001, CSP+LDA/004; --: not evaluated.}}
\label{tab:main_results}
\setlength{\tabcolsep}{2.5pt}
\begin{tabular*}{\textwidth}{@{\extracolsep{\fill}}lccccccc@{}}
\toprule
\textbf{Dataset} & \rev{\textbf{NoAlign EEGNet}} & \rev{\textbf{EA+Safe-BN EEGNet}} & \textbf{EEG-Fusion} & \rev{\textbf{Matched TENT}} & \rev{\textbf{Post-hoc Classical$^\dagger$}} & \rev{\textbf{$\Delta$F1 vs. NoAlign}} & \rev{\textbf{C red. vs. NoAlign}} \\
 & \textbf{F1/C} & \textbf{F1/C} & \textbf{F1/C} & \rev{\textbf{F1/C}} & \textbf{F1/C} & \rev{\textbf{[95\% CI]}} & \rev{\textbf{[95\% CI]}} \\
\midrule
BCI IV-2a local & 0.417/0.541 & 0.523/0.367 & \rev{0.524/0.347} & \rev{\textbf{0.542}/\textbf{0.325}} & 0.456/0.423 & \rev{+0.108 [0.085, 0.129]} & \rev{+0.194 [0.121, 0.271]} \\
BNCI2014-001 & 0.314/0.643 & 0.435/0.518 & \rev{0.480/0.419} & \rev{\textbf{0.582}/\textbf{0.317}} & 0.456/0.330 & \rev{+0.166 [0.111, 0.221]} & \rev{+0.224 [0.116, 0.341]} \\
BNCI2014-004 & 0.607/0.761 & 0.681/0.676 & 0.708/0.592 & \rev{\textbf{0.759}/0.547} & 0.691/\textbf{0.524} & \rev{+0.100 [0.053, 0.150]} & \rev{+0.169 [0.087, 0.249]} \\
\rev{Cho2017 full52} & \rev{0.629/0.719} & \rev{0.708/0.550} & \rev{\textbf{0.709}/\textbf{0.536}} & \rev{0.608/0.547} & \rev{--} & \rev{+0.080 [0.051, 0.110]} & \rev{+0.183 [0.149, 0.216]} \\
\bottomrule
\end{tabular*}
\end{table*}

\subsection{What Actually Drives the Gain?}
\rev{The decomposition is clear: alignment/BN-TTA supplies most F1 gain; routing manages residual reliability. On local IV-2a, NoAlign EEGNet improves from 0.417/0.541 F1/collapse to 0.523/0.367 before routing. Across the core protocols, routing then adds 0.001--0.046 F1 and reduces collapse by 0.020--0.101. Its largest F1 increment is on BNCI2014-001, whereas the IV-2a increment is small. Fallback activates only on BNCI2014-001/004. CSP/LDA retains the lowest BNCI2014-004 collapse, showing why classical models remain informative comparators. EEG-Fusion is therefore a reliability layer above adaptation.}

\subsection{Routing Behavior and Failure Cases}
\begin{figure*}[t]
\centering
\includegraphics[width=\textwidth]{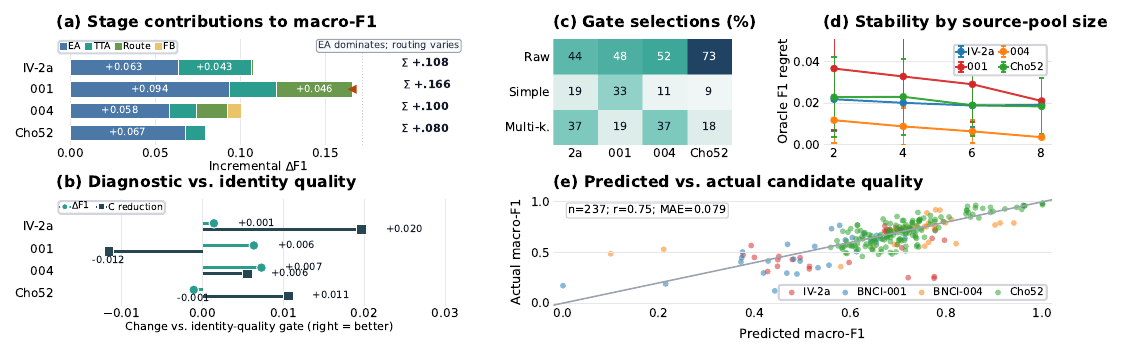}
\caption{\rev{Routing audit. (a) Totals use unrounded values. (b) Identity-quality retains observed-collapse control, isolating richer diagnostics. (c) Selection rates. (d) Mean regret; bars show mean within-target subset SD. (e) Predicted candidate F1 remains imperfect.}}
\label{fig:failure_routing}
\end{figure*}

\rev{Fig.~\ref{fig:failure_routing} summarizes 237 decisions. Replacing diagnostic F1 prediction with identity while retaining observed-collapse control shifts F1 by --0.001 to +0.007 and collapse by --0.012 to +0.020 (Cho52 reduction: 0.011). Thus, source-global quality and observed collapse explain much behavior. Richer diagnostics are modest, dataset-dependent refinements. A three-way replay finds the largest observed-collapse contribution on BNCI2014-004 (+0.039 F1; 0.078 collapse reduction). Regret begins to plateau from 2 to 6--8 source subjects as subset dispersion contracts. Selection remains imperfect: predicted quality correlates with post-hoc F1 ($r=0.75$, MAE 0.079), oracle top-1 is 44\%--74\%, and regret is 0.004--0.021. One BNCI2014-001 case selects EEGNet at 0.560 F1 versus SimpleConv at 0.706.}

\subsection{\rev{Adaptation and Robustness Checks}}
\begin{wraptable}[8]{r}{0.35\columnwidth}
\vspace{-1pt}
\centering
\scriptsize
\caption{\rev{IV-2a Q sweep.}}
\label{tab:q_tradeoff}
\setlength{\tabcolsep}{1.4pt}
\begin{tabular}{@{}ccccc@{}}
\toprule
\textbf{Q} & \textbf{F1} & \textbf{C} & \rev{\textbf{$\Delta$F1}} & \rev{\textbf{s}} \\
\midrule
3 & 0.493 & 0.438 & \rev{--0.030} & \rev{0.68} \\
5 & 0.505 & 0.405 & \rev{--0.018} & \rev{0.82} \\
\textbf{10} & \textbf{0.523} & \textbf{0.367} & \rev{0.000} & \rev{1.10} \\
15 & 0.528 & 0.342 & \rev{+0.005} & \rev{1.41} \\
\bottomrule
\end{tabular}
\vspace{-8pt}
\end{wraptable}
\rev{Table~\ref{tab:q_tradeoff} isolates the IV-2a EA+Safe-BN step budget; ``s'' is single-candidate target time. Q3/Q5 are faster but less reliable. Q15 costs 0.31 s more than Q10 for only +0.005 F1 [--0.008, 0.016] ($p=0.419$), making Q10 a practical elbow. Matched TENT has higher mean core F1 but is 0.100 lower on Cho52; these are secondary paired effects. No tested policy uniformly dominates. An exploratory six-candidate gate selects TENT in 52\%--70\% of core decisions and 0\% on Cho52. Stronger source-global choice on BNCI2014-001/004 motivates better policy calibration. Across IV-2a, BNCI2014-001/004, and Cho9, a fixed $3^4$ routing grid spans at most 0.014 F1, 0.040 collapse, and 0.014 regret. Safe and naive BN-TTA have similar IV-2a gains (0.043 vs. 0.044 F1; $p=0.820$), so safety is a rollback guardrail, not a standalone accuracy claim. At 0.85 imposed majority fraction, F1 falls to 0.593 on BNCI2014-004 and 0.533 on Cho9, requiring prior-aware thresholds. Adaptation dominates target-time cost; deployment remains batch-transductive.}


\subsection{\rev{Limitations and Deployment Scope}}
\rev{The two IV-2a evaluation paths are closely related, and the three core protocols contain only nine held-out subjects. Consequently, the full 52-subject Cho2017 analysis provides the strongest external evidence. Collapse is informative under known or approximately balanced class priors but cannot distinguish model failure from legitimate target imbalance, while probability-based diagnostics may be affected by calibration shift. Evaluation currently includes one matched source-free TTA comparator (TENT); cross-family routing, controlled T-TIME/SPDIM comparisons, cross-dataset gate transfer, and foundation-model integration remain important extensions.}
\section{Conclusion}
\label{sec:conclusion}
This paper presented EEG-Fusion, a failure-informed source-free MI-EEG framework \rev{that predicts candidate quality, observes collapse, and routes accordingly}. It combines alignment, normalization-only adaptation, and source-held-out routing among three neural architectures; classical/physiological models remain comparators. Across LOSO evaluations, EEG-Fusion \rev{yields higher mean macro-F1 and lower collapse than} NoAlign EEGNet. \rev{On full Cho52, routing preserves EA+Safe-BN F1 while significantly reducing collapse. Alignment/TTA supplies most gain. Matched TENT's dataset-dependent rank reversal shows that no tested policy uniformly dominates. EEG-Fusion thus contributes an auditable reliability layer above adaptation. Under known class balance, prediction-stream diagnostics expose failures hidden by aggregate validation; unknown priors and calibrated policy banks remain important steps toward online deployment.}

\def\IEEEbibitemsep{0.7pt}
\bibliographystyle{IEEEtran}
\bibliography{cite}

\end{document}